\documentclass[floatfix,twocolumn,showpacs]{revtex4}
\usepackage{amsmath}
\usepackage{graphicx}
\usepackage[pdftex,colorlinks,citecolor=blue,linkcolor=red,urlcolor=blue]{hyperref}

\usepackage{bm}
\usepackage{bbm}

\newcommand{\ketbra}[2]{\mbox{$|#1\rangle\langle #2|$}}

\newcommand{\ket}[1]{\vert#1\rangle}
\newcommand{\bra}[1]{\langle#1\vert}

\begin{document}

\title{  Extended Techniques for Feedback Control of A Single Qubit}

\author{Y. Yang$^1$, X. Y. Zhang$^1$, J. Ma$^{1,2}$, and X. X. Yi$^1$}
\affiliation{$^1$School of Physics and Optoelectronic Technology,
Dalian University of Technology, Dalian 116024, China\\
$^2$School of Medical Devices, Shenyang Pharmaceutical University,
Shenyang 110016, China}

\date{\today}

\begin{abstract}
The protection of quantum states is challenging  for non-orthogonal
states especially in the presence of noises. The recent research
breakthrough  shows that  this difficulty can be overcome by
feedback control with weak measurements. However, the
 state-protection schemes proposed recently work optimally   only for
special quantum  states. In this paper, \textbf{by applying
different weak measurements, we extend the idea of the
state-protection scheme to protect general states.} We calculate
numerically the optimal parameters and discuss the performance of
the  scheme. Comparison between this extended scheme and the earlier
scheme is also presented.
\end{abstract}


\maketitle

In classical physics, it is possible in principle to  acquire all
information about the state of a classical system by precise
measurements. Namely, the state of a single classical system can be
precisely determined by measurements. This ensures the
measurement-based classical feedback control and makes the feedback
control  beneficial to the manipulation  of classical system.

For a quantum system, however, this is not  possible: If the system
is prepared in one of several non-orthogonal states, no measurement
can determine determinately  which state the system is really in.
Furthermore, Heisenberg's uncertainty principle imposes a
fundamental limit on the amount of information  obtained from a
quantum system, and  the act of measurement necessarily disturbs the
quantum system \cite{1,2,3,4,5} in an unpredictable way. This means
when extend  the measurement-based classical control theory to
quantum system, we need careful examinations of the control scheme.
The extension of the classical feedback  to quantum systems can be
used not only in quantum control \cite{6,7,8,9,10,11}, but also in
quantum information processing, for example in the quantum key
distribution \cite{12} and quantum computing,  as well as in other
practical quantum technologies \cite{13}.

Recent works in this field \cite{14,15,16}  suggested  that we can
balance the information gain from a measurement and the disturbance
caused by the measurement via weak measurement. To be specific, in
Ref.\cite{15} Branczyk {\it et al.} investigated the use of
measurement and feedback control to protect the state of a qubit.
The qubit is prepared in one of two non-orthogonal states \textbf{
 in the $x-z$ plane of the Bloch sphere  and subjected to
noise.} The authors shown that, in order to optimize the performance
of the state protection, one must use non-projective measurements
 to balance the trade-off between information gain
and disturbance. \textbf{The measurement operators used in\cite{15}
are among the $y-$ axis and the subsequent correction is a rotation
about the $z-$axis}. This scheme was realized recently \cite{14},
where the stabilization of non-orthogonal states of a qubit against
dephasing was experimentally reported. It is shown that the  quantum
measurements applied in the experiment play an important role in the
feedback control. \textbf{We should notice that the measurements
used in\cite{14} are different to those in \cite{15}, namely, its
measurement operators  are along the $z-$axis and the correction is
about the $y-$ axis. Geometrically, for initial states in  the $x-z$
plane, the dephasing noise can not map the initial states out of the
$x-z$ plane, then all states including the initial states, the
states passed the noise and measurements as well as  the final stats
are in the $x-z$ plane in\cite{14}, this is the difference between
\cite{14} and \cite{15} from the geometric viewpoint. We will modify
the measurement operators in \cite{14} and use it in this paper.}

{With these knowledge in quantum information science
\cite{17,18,19}, one may wonder if the weak measurement used  in the
scheme is also the best one for the protection of general states?
I.e., $M_+ =\cos(\chi/2)|0\rangle\langle 0| +
\sin(\chi/2)|1\rangle\langle1|\,, $ and $M_- =
\sin(\chi/2)|0\rangle\langle 0| +\cos(\chi/2)|1\rangle\langle 1|\,,$
are these measurements best for the protection of general states?
Are there other measurements that can better the performance of the
scheme for general states? In this paper, we shall shed light on
this issue by introducing   different measurements for the feedback
control. We find  that the scheme can be extended to protect general
quantum states with the new weak measurement. We derive the
performance and give the parameters best for the performance, a
discussion on this extended  scheme is also presented.}

Consider two non-orthogonal states that we want to protect from noise,
\begin{equation}
\label{initial state} \ket{\psi_{\pm}}{=}
\cos\frac{\theta}{2}\ket{+}{\pm}e^{i\phi}\sin\frac{\theta}{2}\ket{-},
\end{equation}
with  $\ket{\pm}{=}\frac{1}{\sqrt{2}}(\ket{0}\pm\ket{1})$, the
corresponding density matrices are given by
$\rho_\pm{=}\ketbra{\psi_\pm}{\psi_\pm}$. Note that $|\psi_+\rangle$
and $|\psi_-\rangle$ are non-orthogonal and are more general than
the states in \cite{14,15}, the overlapping of the two states is
independent of $\phi$, but depends on $\theta$,
$\langle\psi_+|\psi_-\rangle=\cos\theta.$  In fact,
$\ket{\psi_{\pm}}$ are rotated about the $x$-axis with respect to
the Branczyk's one, this may offer a chance to improve the fidelity
given by the previous proposals \cite{14,15} for general states of a
qubit.

The qubit is subjected to  dephasing noises \cite{14,15}. We shall
use $\{\ket{0},\ket{1}\}$ as the basis of the qubit Hilbert space,
and define the Pauli operator $Z$ as $Z\ket{0}{=}\ket{0},
Z\ket{1}{=}{-}\ket{1}$, similar definitions are for Pauli matrices
$X$ and $Y$. The dephasing noise can be described by a phase flip
$Z$ with probability $p$ and with probability $1-p$ that the system
remains unchanged. The density matrix of the qubit passed  through
the noisy channel is,
\begin{equation}
\rho^{'}_{\pm}= (1-p) \rho_{\pm}+p Z \rho_{\pm} Z.\label{noise}
\end{equation}

The purpose of this paper  is to find  better measurements and
controls to send the qubit back as close as possible to its initial
state. For this purpose, we use a quantum operation $\mathcal{C}$ as
a map acting on the single qubit to describe the controls and
measurements,
$$\mathcal{C}(\rho^{\prime})=Y_{+\eta}M^{\prime}_{+}
\rho^{\prime}M_{+}^{\prime\dagger}Y_{+\eta}^{\dagger}+Y_{-\eta}M^{\prime}_{-}
\rho^{\prime}M_{-}^{\prime\dagger}Y_{-\eta}^{\dagger}.$$  The
notations of $Y$ and $M^{\prime}$ will be given later. To quantify
the performance of $\mathcal{C}$, we use the average fidelity
$\overline{F}$ \cite{14,15} between the noiseless input state and
the corrected output state as a measure,
\begin{align}
\overline{F}&=\tfrac{1}{2}[\bra{\psi_+}\mathcal{C}(\rho'_+)\ket{\psi_+}
+\bra{\psi_-}\mathcal{C}(\rho'_-)\ket{\psi_-}]\nonumber\\
&=\tfrac{1}{2}(F_{\psi_+}+F_{\psi_-})\,.
\end{align}
This measure quantifies the performance well when $\ket{\psi_+}$ and
$\ket{\psi_{-}}$ are sent into the control with equal probability.
\begin{figure}
\includegraphics*[width=8cm,height=4cm]{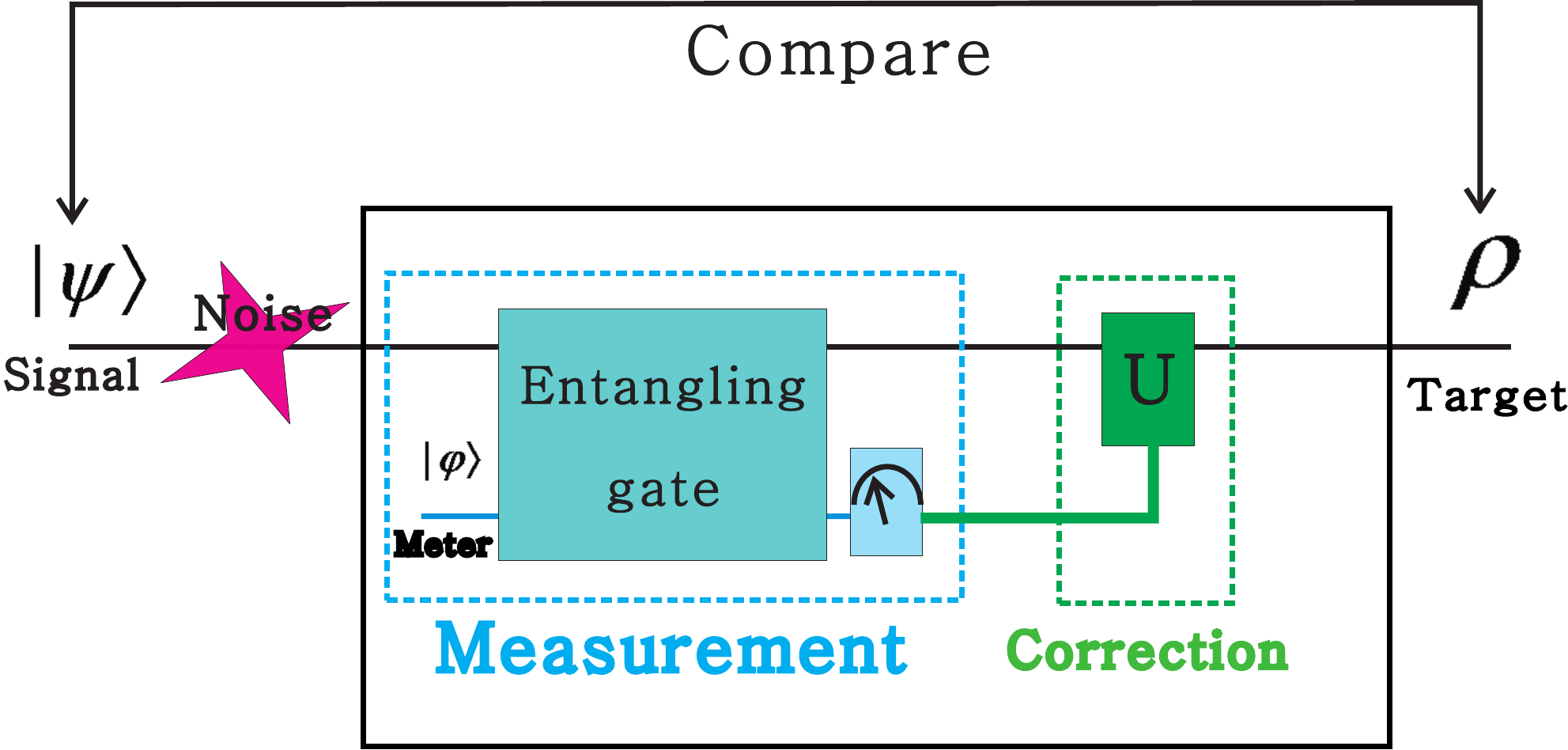}
\caption{Illustration of the scheme. The meter qubit  was entangled
with the qubit (for protection)   which has  passed through the
noise channel. After the measurement and correction, we can get a
signal. Then we compare the signal with  the initial state and apply
a feedback control to the qubit. An average fidelity is define and
used to determine the parameters in the  feedback control. In the
earlier scheme, the authors use a
$\ket{\varphi}{=}\cos\frac{\chi}{2}\ket{+}{+}\sin\frac{\chi}{2}\ket{-}$
as the meter-qubit state, while in the present scheme a complex
phase factor is introduced, i.e. the meter state is,
$\ket{\varphi}{=}\cos\frac{\chi}{2}\ket{+}{+}e^{i\beta}
\sin\frac{\chi}{2}\ket{-}.$}\label{FIG:1}
\end{figure}

To  find  a good  control procedure, we must first find the
appropriate measurement which has to have the following two
features. First, it must be a weak measurement, that is, it can not
 completely disturb  the system. Second, it has to be
strength-dependent, such that we can adjust the strength of the
measurement as we need. This family of weak measurements in the
logical basis $\{\ket{0},\ket{1}\}$ can be written as,
\begin{align}\label{eq:M'_{+}}
M'_+ &= \cos(\chi/2)|0\rangle\langle 0| +  e^{i\beta}\sin(\chi/2)|1\rangle\langle 1|\,, \\
\label{eq:M-} M'_- &= e^{i\beta}\sin(\chi/2)|0\rangle\langle 0| +
\cos(\chi/2)|1\rangle\langle 1|\,.
\end{align}
In contrast  to the measurements used in Ref.\cite{14,15,16}, $M_+
=\cos(\chi/2)|0\rangle\langle 0| + \sin(\chi/2)|1\rangle\langle1|\,,
$ $M_- = \sin(\chi/2)|0\rangle\langle 0|
+\cos(\chi/2)|1\rangle\langle
 1|\,,$ a new  parameter $\beta$ was introduced in this
weak measurement \cite{20,21}. Here $\chi$  ranges from 0 to $\pi/2$
\cite{20}, we can change the value of the parameter $\chi$  to
adjust  the strength of measurement. The corresponding positive
measurement operators are given by $\Pi_\pm{=}M^{'\dag}_\pm M'_\pm
{=}[\mathbbm{1}\pm\cos{(\chi)}Z]/2$, with $\mathbbm{1}$ being the
identity operator. Clearly, $\chi=0$ describes the projective
measurement, while $\chi=\frac{\pi}{2}$, do nothing. At first
glance, this proposal is trivial, i.e., the initial states (the
state sent into protection) are rotated about $x-$axis in the Bloch
sphere with respect to that in Ref.\cite{15},  by properly choosing
$\beta$,  the next measurements $M_+^{\prime}$ and $M_-^{\prime}$
may send them back, then the resulting states will return to that in
the earlier proposal, and the performance can not be improved. We
will show later that this is not the case.

Our main task is to figure out how the  parameter $\beta$  affects
the results of the control, and  if the parameter $\beta$ can better
the performance. The correction performed in this paper is the same
as that in \cite{14}, i.e.,  $Y_{\pm\eta} = \exp(\pm
i\tfrac{1}{2}\eta Y)$ representing a rotation   with an angle $\eta$
around the $y-$axis of the Bloch sphere. All parameters should be
optimized for the performance of the control.

Straightforward calculation show that the average fidelity of the
control is a function of  $\theta, \phi, \eta, \chi, \beta$ and $p$,
\begin{multline}
\label{fertility}
\overline{F'}(\theta,p,\chi,\eta,\phi,\beta)=\\
\tfrac{1}{2}\left[1+\cos\theta\cos\chi\sin\eta+
\cos\eta\cos^{2}\phi\sin^{2}\theta\right.\\
+\tfrac{(1-2p)}{2}\sin\chi\left(2\cos\beta
(\cos\eta\cos^{2}\theta+\sin^{2}\theta\sin^{2}\phi)\right.\\
\left.\left.-\sin\beta\sin\eta\sin^{2}\theta\sin2\phi\right) \right]
\,.
\end{multline}
  {For each $\theta,$ $\phi$ and $p$, there are an optimum
measurement strength $\chi,$ correction angle $\eta,$ and
measurement parameter  $\beta$, which maximizes the average
fidelity. First we start with  $\eta$.} The  $\eta$ which optimizes
the average fidelity can be given by,
\begin{multline}
\label{N}
\eta_{\rm opt}(\theta, p, \chi,\phi,\beta)\\
=\arctan{\frac{\cos\theta\cos\chi-\tfrac{1}{2}(1-2p)
\sin\beta\sin^{2}\theta\sin2\phi\sin\chi}{\cos^{2}\phi\sin^{2}\theta
+(1-2p)\cos^{2}\theta\sin\chi\cos\beta}
} \,.\\
\end{multline}
  {Substituting the optimum $\eta_{opt}$ into the average
fidelity, we have,}
\begin{multline}
\label{fertility'}
\overline{F'}(\theta,p,\chi,\phi,\beta)=
\tfrac{1}{2}+\tfrac{1}{2}(1-2p)\cos\beta\sin^{2}\theta\sin^{2}\phi\sin\chi\\
+\frac{1}{2}\left[(\cos\theta\cos\chi
-\tfrac{1}{2}(1-2p)\sin\beta\sin^{2}\theta\sin2\phi\sin\chi)^{2}\right.\\
+(\cos^{2}\phi\sin^{2}\theta
\left.+(1-2p)\cos^{2}\theta\sin\chi\cos\beta)^{2}\right
]^{\tfrac{1}{2}}\,.
\end{multline}
We can see that when $\phi=0$,
$\overline{F'}(\theta,p,\chi,\phi,\beta)$ reduces to
\begin{eqnarray}
&\ &\overline{F'}|_{\phi=0}=\frac 1 2 +\frac 12\left
[\cos^2\theta\sin^2\chi\right. \nonumber\\
&\
&\left.+(\sin^2\theta+(1-2p)\cos^2\theta\sin\chi\cos\beta)^2\right
]^{\frac 12}.
\end{eqnarray}
Obviously, $\beta=0$ maximize the average fidelity $\overline{F'}$,
this is exactly the case discussed in Ref. \cite{14,15}. So, for the
initial states lying in the $xz-$ plane of the Bloch sphere, the
weak measurements with $\beta=0$ already maximize the performance.

\begin{figure}
\includegraphics*[width=8.5cm,height=6cm]{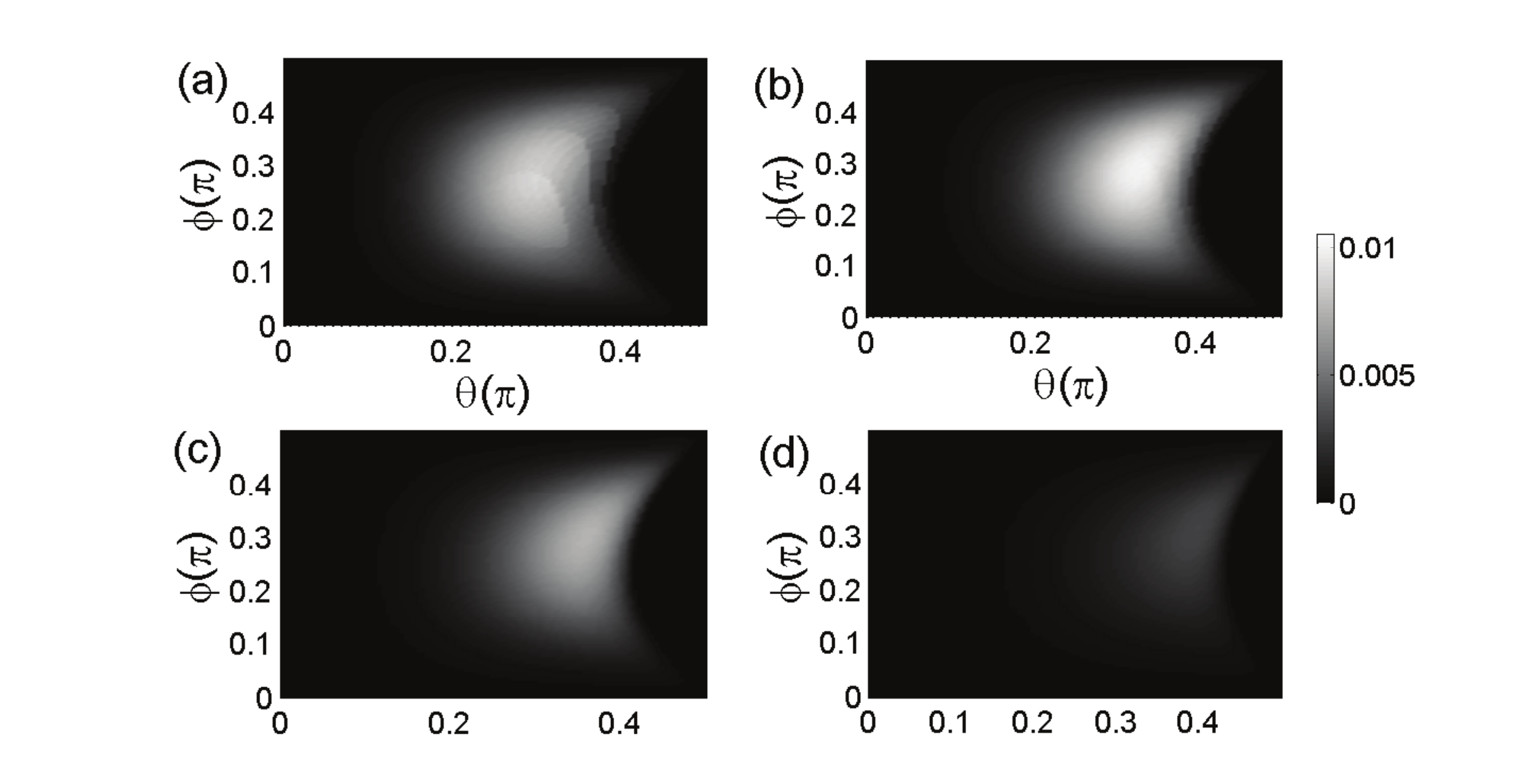}
\caption{This figure shows how much our scheme improve  the
performance of the state protection for general qubit states. The
improvement is quantified by $\delta_{F}$, which  is plotted as a
function of $\theta$ and $\phi$. For different  $p$, the improvement
is different, as (a), (b), (c) and (d) show.  (a)$p=0.10;$
(b)$p=0.20;$ (c)$p=0.30;$ (d)$p=0.40.$}\label{FIG:2}
\end{figure}
To find the optimal feedback control for $\phi\neq 0$,  we  follow
the procedure in \cite{14}. Here again $\theta$ and $\phi$ are
related to the initial state of the qubit, while $p$ characterizes
the noise and  is regarded as a fixed value, $\chi$ and $\beta$ are
related to the measurement procedure, $\eta$ denotes the correction
parameter.
\begin{figure}
\includegraphics*[width=8.5cm,height=6cm]{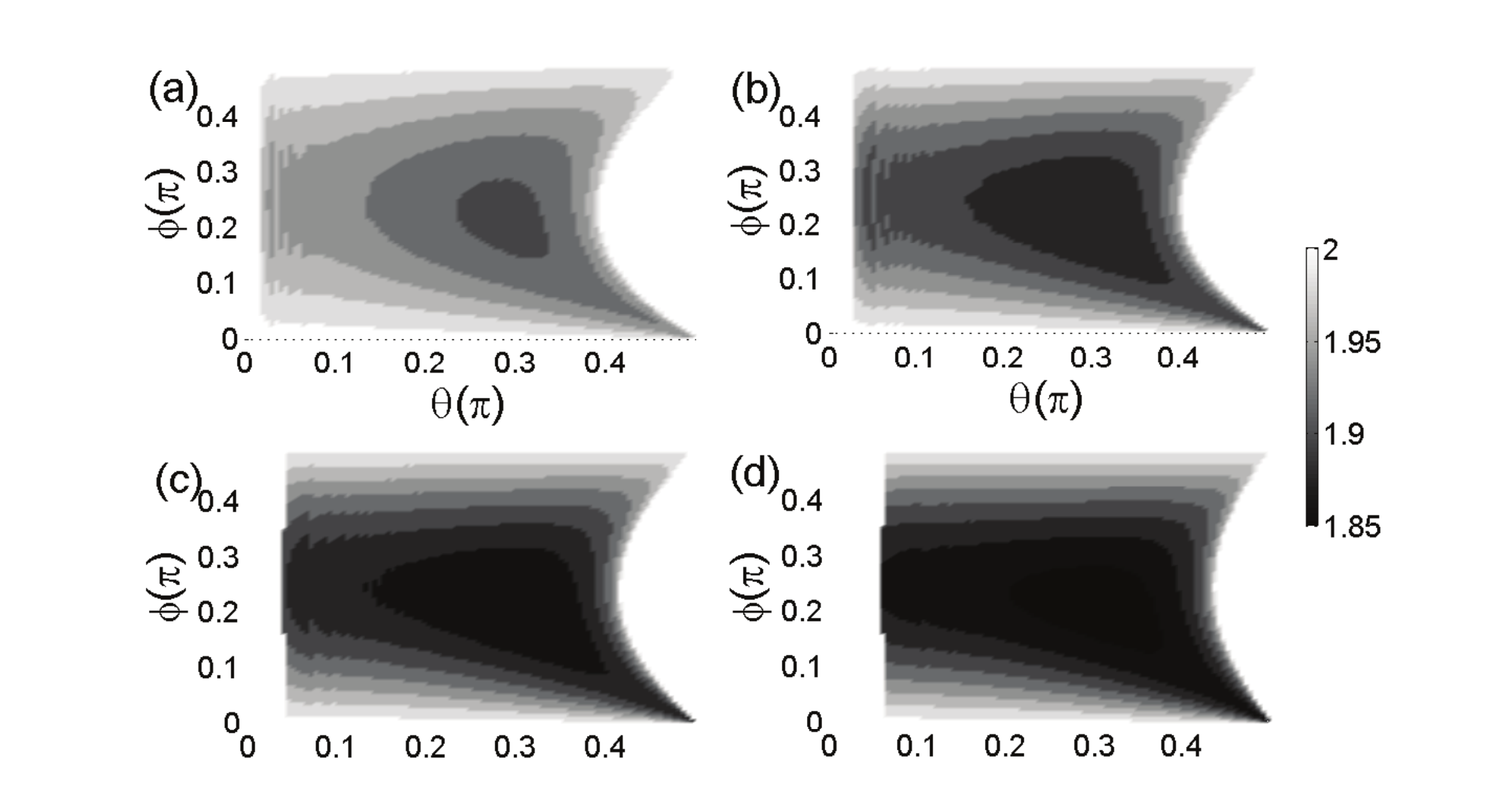}
\caption{The $\beta$ which maximizes $\delta_{F}$ as a function of
$\theta$ and $\phi$ for different $p$, (a)$p=0.10;$ (b)$p=0.20;$
(c)$p=0.30;$ (d)$p=0.40.$}\label{FIG:3}
\end{figure}
By the same procedure as in the earlier works, we maximize the
fidelity of the control over the remaining parameters $\chi$,
$\theta, \phi, \beta$ and $p$. The analytical expression for the
fidelity is complicated, so we choose to find the optimal parameters
by numerical simulations. As aforementioned, we have already had the
relations between the average fidelity and the initial parameters
$\theta$ and $\phi$. We shall use $\delta_{F}{=}F'_{opt}{-}F_{opt}$
to quantify the improved fidelity due to the parameter $\beta$,
select results are presented in Fig.\ref{FIG:2}, where $F'_{opt}$
denotes the optimal fidelity in our paper, while  $F_{opt}$ denotes
that by the scheme in Ref.\cite{14,15}, i.e., with $\beta=0.$ The
optimized $\beta$ would depend on $\theta$ and $\phi$ and is shown
in Fig.\ref{FIG:3}.
\begin{figure}
\includegraphics*[width=8cm,height=5cm]{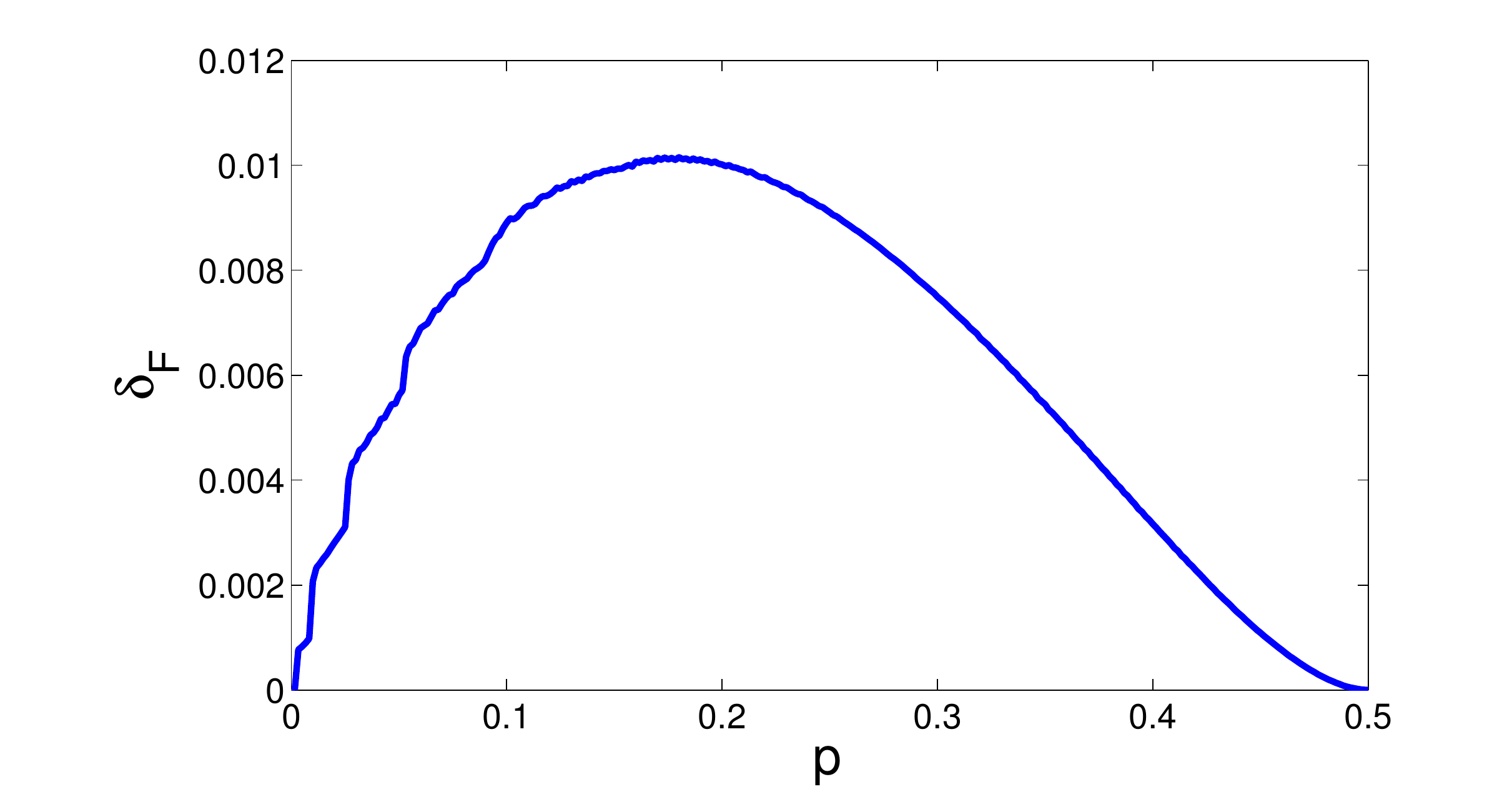}
\caption{The fidelity difference $\delta_{F}$ versus $p$.}
\label{FIG:4}
\end{figure}
Fig.\ref{FIG:2} plots the improvement of the average fidelity as a
function of the original states (characterized by $\theta$ and
$\phi$) with different amount of noise (characterized by $p$). We
note that there are no improvement  for the following cases. If
$p=0$, there is no noise and so the state is not perturbed, in this
case the  fidelity is 1 for all original states including $\phi=0$
and the  measurement strength is $\chi=\frac{\pi}{2}$ (do nothing).
When $\theta=\frac{\pi}{2}$, the state $|\psi_+\rangle$ and
$|\psi_-\rangle$ are orthogonal, the earlier scheme gives unit
fidelity, hence there is no room to improve the performance. When
$\theta=0$ the two states are equal and point along the $x-$axis,
these states are also the same as that in the earlier scheme,
leading to zero improvement. If $\phi=\frac{\pi}{2}$, nothing should
change since the two states would interchange by this control.
Finally, when $\phi=0,$ the initial states return to the earlier
scheme. Fig.\ref{FIG:3} shows the parameter $\beta$, which maximize
the average fidelity as a function of the original states and the
amount of noise $p$. As expected, non-zero maximal $\beta_{opt}$
exists. To show clearly the dependence of the improvement on the
noise strength, we plot $\delta_F$ in Fig. \ref{FIG:4} as a function
of $p$. The maximal improvement arrived at about  p=0.1800, the
corresponding improvement is $\delta_{F}$=0.0102.

\begin{figure}
\includegraphics*[width=6.5cm,height=5.5cm]{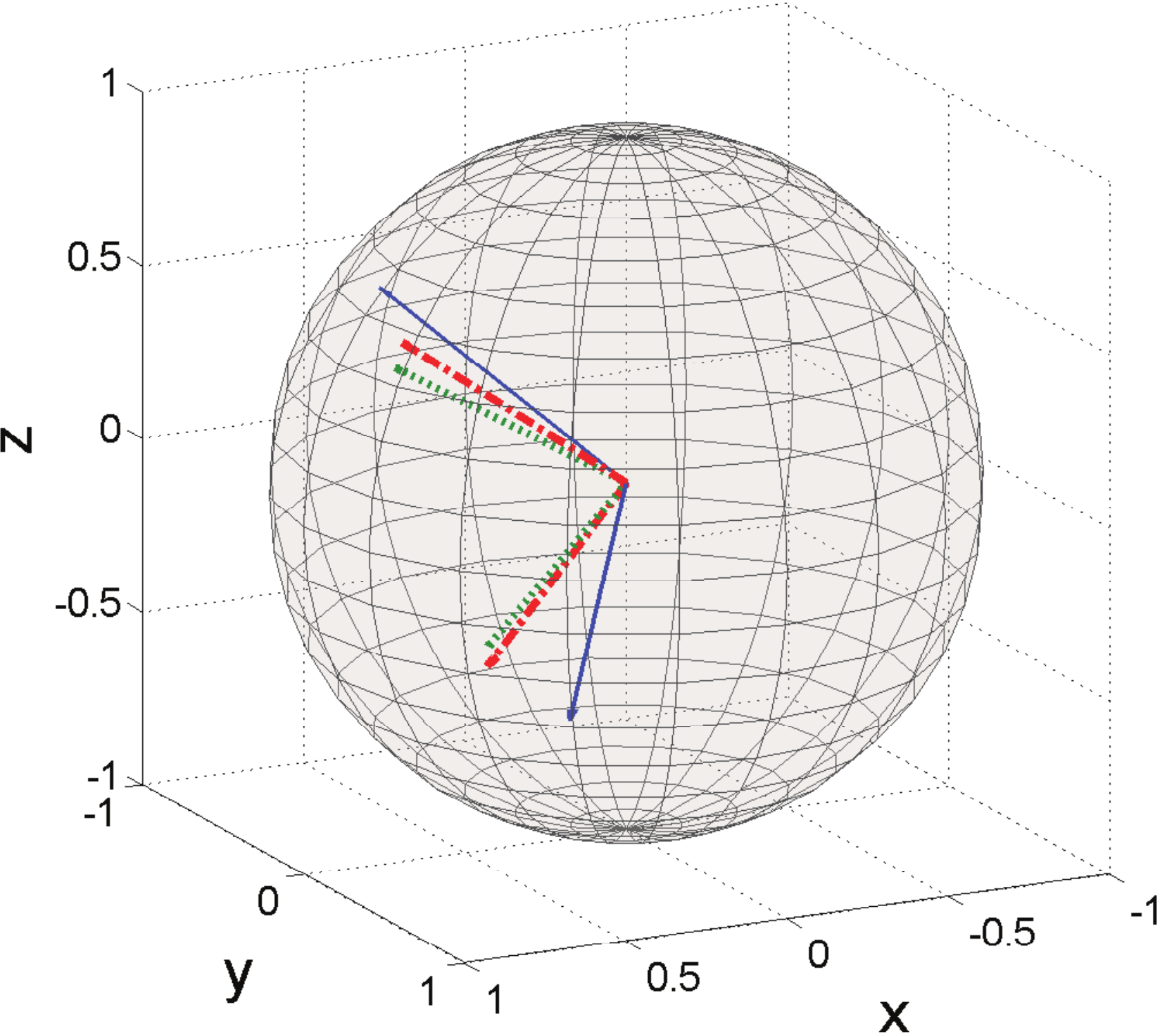}
\caption{Bloch vectors of the original states(blue-solid), the
resulting states by our scheme (red-dashed)  and the resulting
states in \cite{15} (green-dotted). The parameters chosen are,
$p=0.18, \eta=0.7913, \chi=0.8583, \beta=5.8905, \theta=1.0155,
\phi=0.8976.$ All parameters except $p$ are in units of arc. }
\label{FIG:7}
\end{figure}

{For developing an intuitive picture, we now take a snapshot for the
states going through the control and measurement. Suppose the
initial state is $|\psi_+\rangle$ with $\phi=\frac{\pi}{4}$, and let
$\{|0\rangle, |1\rangle\}$ be a basis for the Hilbert space. In
terms of density matrix, the initial state is
$$\rho_{+}=\frac 1 2 \mathbbm{1}+\frac 1 2\cos\theta\cdot\sigma_{x}
-\frac{\sqrt{2}}{4}\sin\theta\cdot\sigma_{y}+\frac{\sqrt{2}}{4}\sin\theta\cdot\sigma_{z},$$
This state lies in the $z=-y$ plane and points along the direction
with an angle $\theta$ from the $x-$axis.  The state  passed the
noisy channel is $\rho^{'}_{+},$
\begin{eqnarray}
\rho^{'}_{+}&=&\frac 1 2\mathbbm{1}\nonumber\\
&+&(\frac 1 2-p)\cos\theta\cdot\sigma_{x} \nonumber\\
&+&\frac{\sqrt{2}}{2}(p-\frac 1 2)\sin\theta\cdot\sigma_{y}\nonumber\\
&+&\frac {\sqrt{2}} {4}\sin\theta\cdot\sigma_{z},
\end{eqnarray}
we see that the $z-$component of the Bloch sphere remains unchanged,
while the $x-$ and $y-$ components are shortened by $(1-2p)$ times
due to the noise. The resulting state (unnormalized) immediately
after the measurement is denoted by $\rho^{m}_{+},$ and it takes,
\begin{eqnarray}
\rho^{m}_{+}&=&\frac 1 2
(1+\frac{\sqrt{2}}{2}\cos\chi\sin\theta)\cdot\mathbbm{1}\nonumber\\
&+&\frac 1
2(1-2p)(\cos\beta\cos\theta+\frac{\sqrt{2}}{2}\sin\beta\sin\theta)\sin\chi\cdot\sigma_{x}\nonumber\\
&+&\frac 1 2
(1-2p)(\cos\theta\sin\beta-\frac{\sqrt{2}}{2}\cos\beta\sin\theta)\sin\chi\cdot\sigma_{y}\nonumber\\
&+&\frac 1 2
(\cos\chi+\frac{\sqrt{2}}{2}\sin\theta)\cdot\sigma_{z}.\label{statem}
\end{eqnarray}
Finally after the correction $Y_{+\eta}$, the unnormalized states
has been mapped into,
\begin{widetext}
\begin{eqnarray}
\rho^{c}_{+}&=&\frac 1 2
(1+\frac{\sqrt{2}}{2}\cos\chi\sin\theta)\cdot\mathbbm{1}\nonumber\\
&+&\frac 1 2 \left(\sin\eta(\cos\chi+\frac{\sqrt{2}}{2}\sin\theta)
+(1-2p)\cos\eta\sin\chi(\cos\beta\cos\theta+\frac{\sqrt{2}}{2}\sin\beta\sin\theta)\right )
\cdot\sigma_{x}\nonumber\\
&+&\frac 1 2
(1-2p)(\cos\theta\sin\beta-\frac{\sqrt{2}}{2}\cos\beta\sin\theta)\sin\chi\cdot\sigma_{y}\nonumber\\
&+&\frac 1 2 \left(\cos\eta(\cos\chi+\frac{\sqrt{2}}{2}\sin\theta)+
(-1+2p)\sin\eta\sin\chi(\cos\beta\cos\theta+\frac{\sqrt{2}}{2}\sin\beta\sin\theta)\right
)\cdot\sigma_{z}.
\end{eqnarray}
\end{widetext}
Note that this state is also unnormalized. For a specific set of
$\theta$, $\phi$ and $p$, the resulting state together with the
resulting state  in Ref.\cite{14} are illustrated in Fig.
\ref{FIG:7}. This shows clearly  that our resulting states are more
close to the initial state than that given by the proposal with
$\beta=0$. As shown, the new measurements can do better than the
earlier one for general quantum states. This suggests that we can
apply the new set of measurements to the feedback control. Now we
examine how much this new scheme improves the fidelity with respect
to the schemes with measurements "do nothing" and "strong
measurement" (Helstrom).}

Before processing,  we briefly review the two special cases of the
schemes, which differ from each other at the measurements: In  the
zero strength measurement, $\cos\chi=0$, namely, no measurement is
applied. So the state protection with this measurement is called "do
nothing" (DN) control scheme; The projective measurement is applied
with maximum strength ($\cos\chi=1$), with which the protection
scheme had already been named as "Helstrom" (H) scheme\cite{22}. In
fact, DN control is actually not a measurement-based control because
of no application of measurement to quantum states. And H scheme is
not what we need, because it makes an unnecessary correction to the
system. To quantify the fidelity difference between these schemes,
we define
\begin{equation}
F_{imp}=F_{opt}^{\prime}-max\{F_{DN},F_{H}\}
\end{equation}
as a measure to quantify the difference, where $F_{DN}$ is the
fidelity of DN control scheme, while $F_H$ represents the fidelity
of the H scheme.
\begin{figure}
\includegraphics*[width=8cm,height=5cm]{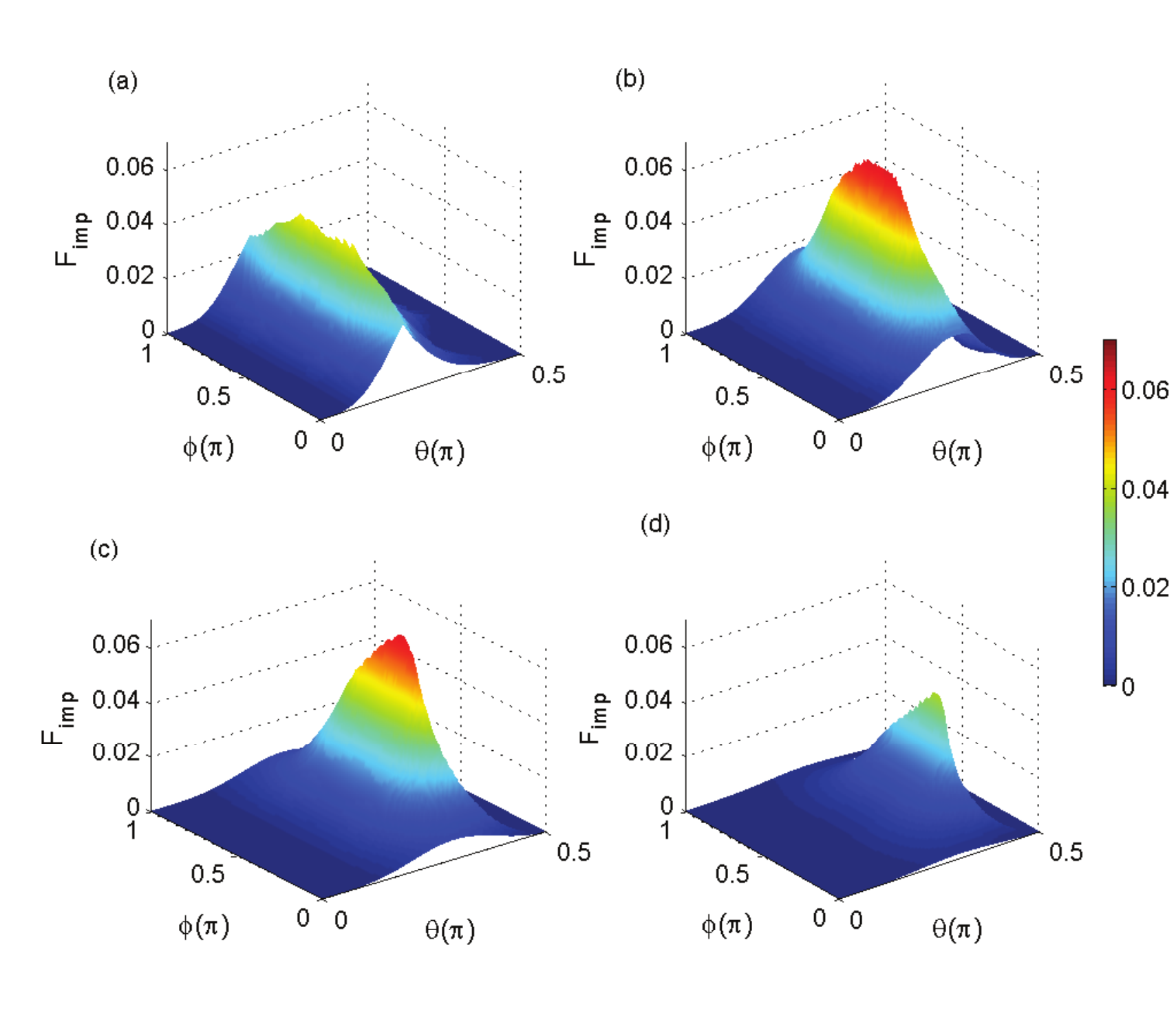}
\caption{$F_{imp}$ versus $\theta$ and $\phi$ with different
$p$,(a)$p=0.10$; (b)$p=0.20$; (c)$p=0.30$; (d)$p=0.40.$ This figure
shows  the improvement of our scheme  over the DN and H
schemes.}\label{FIG:5}
\end{figure}
We have performed numerical simulations for $F_{imp}$, selective
results are presented in Fig.\ref{FIG:5} and Fig. \ref{FIG:6}. In
Fig.\ref{FIG:5}, we present  $F_{imp}$ as a function of  $\theta$
and $\phi$ for different $p$. A common feature is that $F_{imp}$
reach its maximum at around $\theta=\pi/4$ and $\phi=\pi/4$. For
different $p$, the improvement in the fidelity  is different.
\begin{figure}
\includegraphics*[width=8cm,height=5cm]{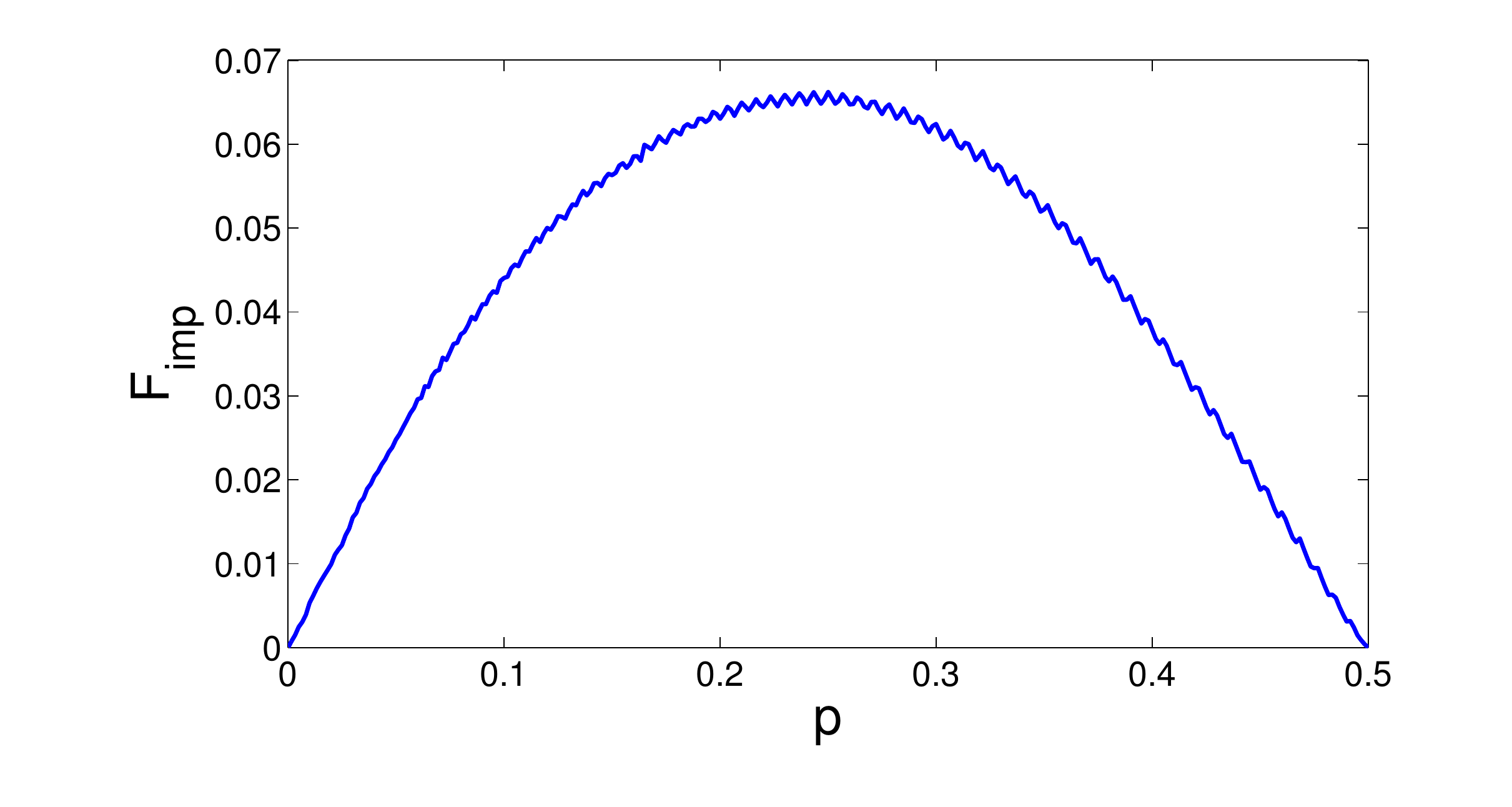}
\caption{$F_{imp}$ as a function of $p$. In this figure, $F_{imp}$
is numerically optimized over $\theta$ and $\phi$ for each $p$. $p$
runs from 0 to 0.5, covering all possible choices.} \label{FIG:6}
\end{figure}

To show the dependence of $F_{imp}$ on $p$ clearly, we plot the
maximum $F_{imp}$ versus the parameter $p$ in Fig. \ref{FIG:6} with
different $\theta$ and $\phi$. As the figure shows,   when
$p=0.2501$, $F_{imp}$ reaches the maximum value 0.0662. Although the
improvement is small, it can  work under most conditions  and it
does improve the state protection over  other schemes with different
measurements\cite{16}. This tells that the scheme without the
parameter $\beta$ is not the best scheme for state protection of
general states.

{It is illustrative to view the difference between our scheme (see
Fig.\ref{fig8}(Right top)) and the scheme (Fig.\ref{fig8}(Left top))
in Ref.\cite{14} on the Bloch sphere. In Fig.\ref{fig8}(Left top),
we can see that the original states $|\psi_+\rangle$ and
$|\psi_-\rangle$ (green) are shorten by the noise, but the
$z-$component of the Bloch vector remains unchanged (pink vector on
the Bloch sphere, i.e., $\rho^{\prime}_{\pm}$). The measurements
lengthen the Bloch vectors(blue, i.e.,
$M_+^{\prime}\rho^{\prime}M_+^{\prime\dagger}$) and diminish the
angle between the Bloch vector and the $z-$ axis. We should remind
that the Bloch vectors remains in the $xz-$ plane in the whole
process of  measurements and controls, this is the core difference
between the scheme in \cite{14} and ours. This difference offers us
a room to improve the performance of the control.}

{In our scheme, the original states are rotated about the $x-$axis
with respect to the earlier scheme, see Fig.\ref{fig8}(Right top).
The effect of the noise is not only  to shorten the length of the
Bloch vector of the states, but also map the Bloch vector out of the
plane of the original states. When the measurement is made, two
things happen, as Fig.\ref{fig8} (Right top) shows. (1) The Bloch
vector is lengthened, in other words, the state become more pure,
see also Eq.(\ref{statem}). (2) The $x$ and $y$ components of the
Bloch vector is mixed, in contrast to the proposal with $\beta=0$.
As a consequence, the next rotation $Y_{\pm\eta}$ about the $y-$axis
may make the resulting states (red vector) more close to the
original states with respect to the earlier scheme.
\begin{figure}
\includegraphics*[width=0.45\columnwidth,
height=0.43\columnwidth]{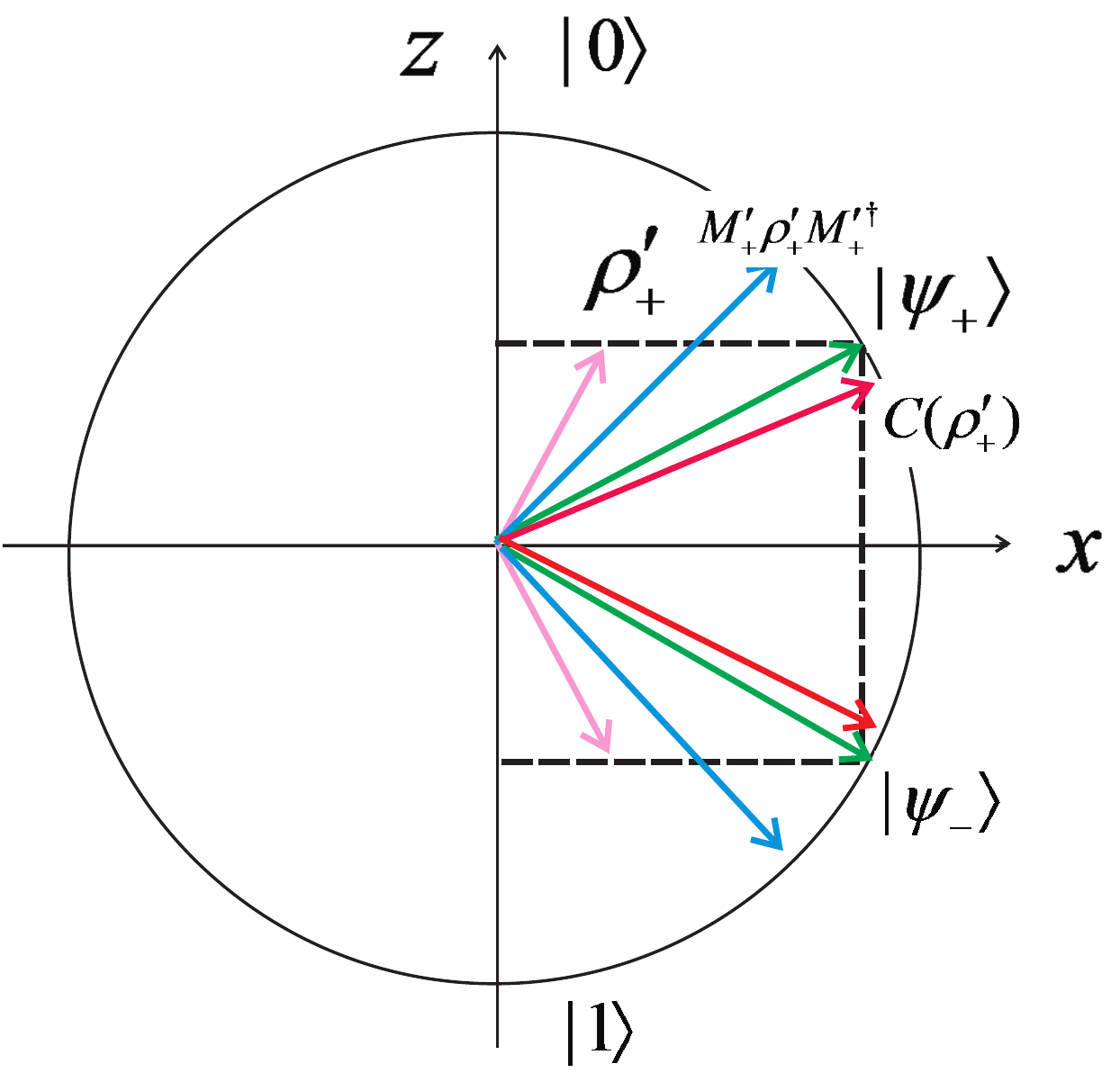}
\includegraphics*[width=0.46\columnwidth,
height=0.43\columnwidth]{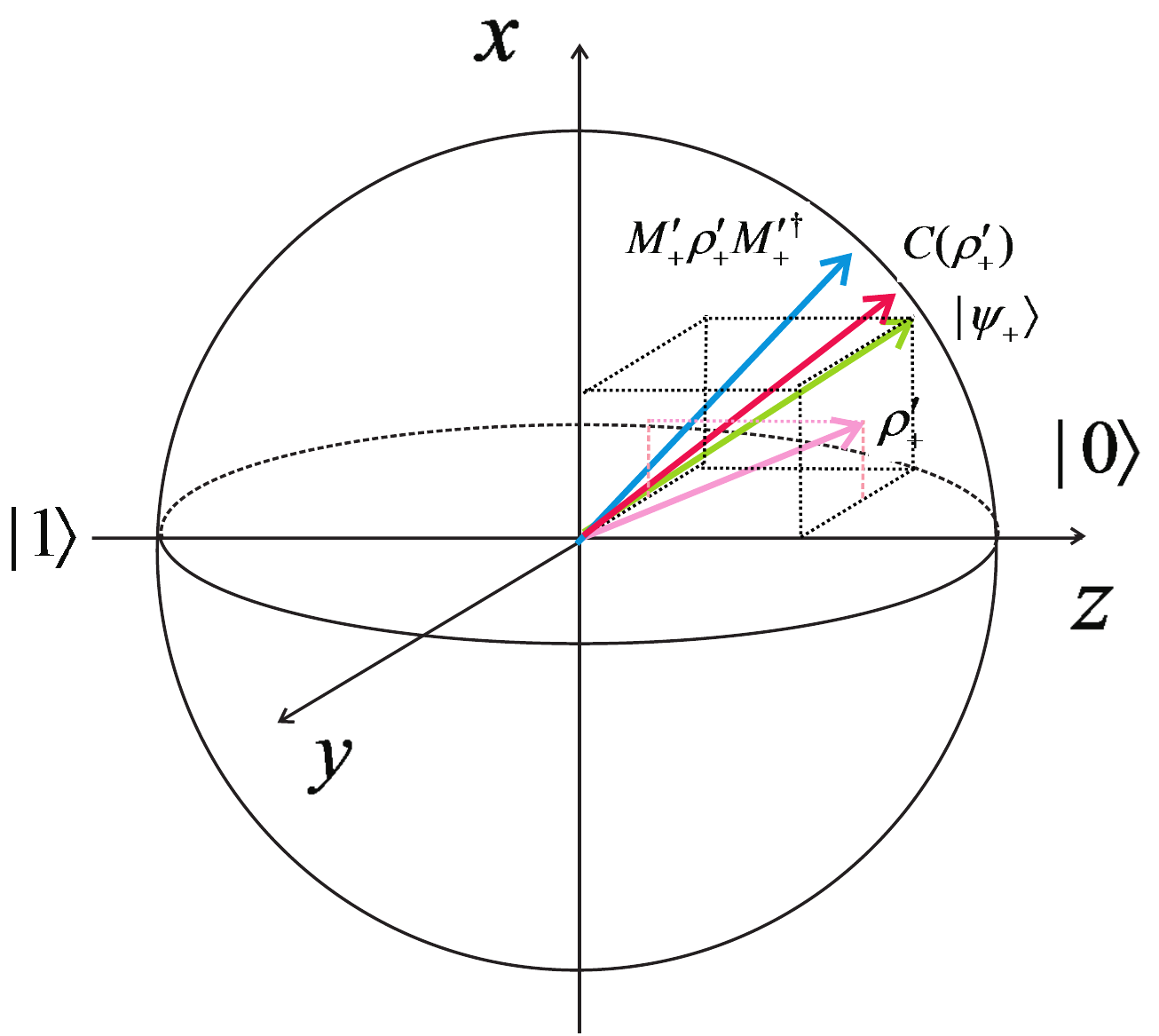}
\includegraphics*[width=0.6\columnwidth,
height=0.5\columnwidth]{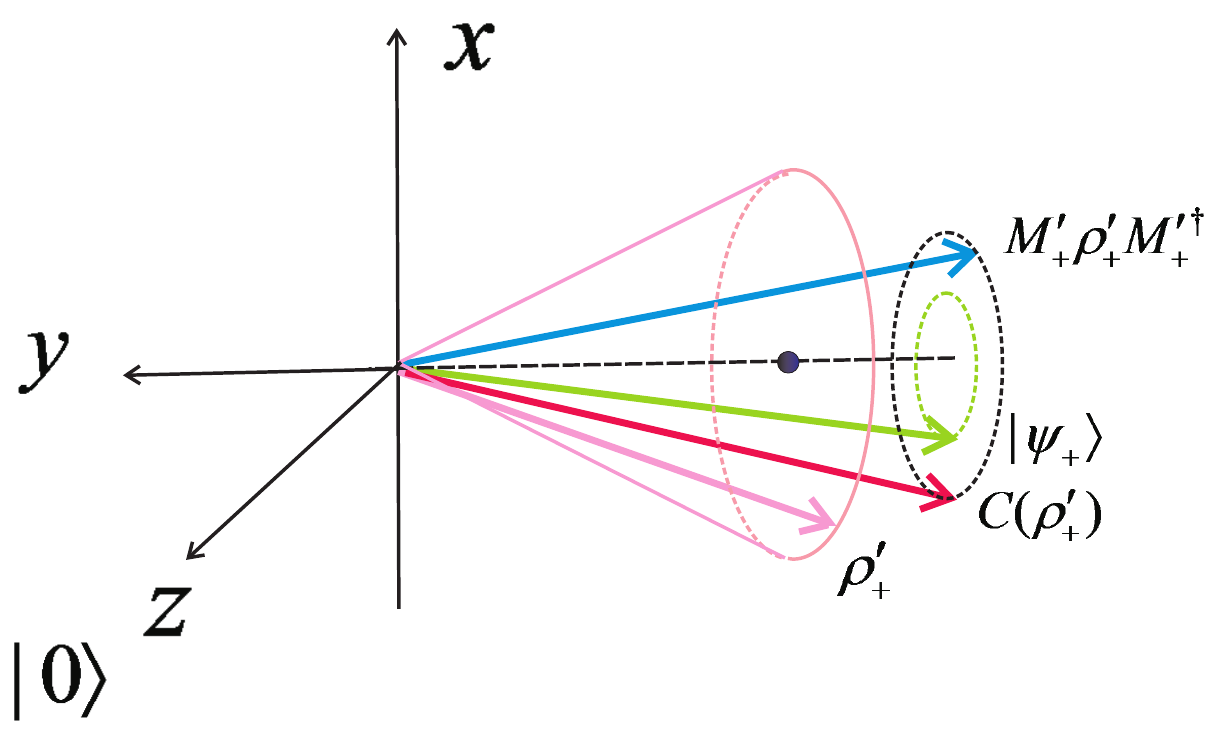} \caption{Bloch sphere
representation of the initial states (green), the states after the
noise (pink), the states after passing the measurement (blue) and
correction (red). Left top is for the scheme in Ref.\cite{14}, while
the Right top figure is for ours. Notice that the Left top figure
shows only the $xz-$plane of the Bloch sphere, and the axis of the
Bloch sphere in the Left and Right top figures are different. {The
Bloch vectors of the initial state, the state after the
noise(post-noise state) and the state after the
measurement(post-measurement state) form three cones, sharing an
axis, i.e., the $y-$axis, which pass perpendicularly through the
centre of the bases(see the bottom figure). The three cones have a
common apes, i.e., the origin of the Bloch sphere. One optimal
scheme is to use measurement operators that map the two post-noise
states to the cone formed by the initial states as close as
possible. The correction is a rotation about the $y-$axis, which
would rotate the post-measurement states to the initial states as
close as possible.}} \label{fig8}
\end{figure}}

Both control schemes in \cite{14} and \cite{15} are optimal for
depolarizing noise and states lying in the $x-z$ plane, the
depolarizing noise keeps these particular states in the $x-z$ plane
and maintains the trace distance between the two states. If the
original states are not  in the $x-z$ plane, the depolarizing noise
can not maintain the trace distance between the two states and
causes the plane in which the two states lie  to rotate as the
states pass through the depolarizing channel. The optimal control
scheme will depend on the orientation of the post-noise states.
\textbf{From the optimality proof in Ref. \cite{15}, we find that
one optimal scheme is to use measurement operators to  prolong the
Bloch vectors of the post-noise states, and the correction is to
bring the post-measurement states to the initial states as close as
possible. The measurements and the correction are closely connected
for a high performance.} In the present scheme, the optimal scheme
is to use measurement operators that can map the two post-noise
states as close as possible to the cone formed by the initial
states. Specifically, the Bloch vectors of the initial state, the
post-noise state and the post-measurement state form three cones(see
the bottom figure of Fig. \ref{fig8}), these cones share an axis:
the $y-$axis, which pass perpendicularly through the centers  of the
bases. The three cones have a common apes, i.e., the origin of the
Bloch sphere. One optimal scheme is to use measurement operators
that map the two post-noise states very close to the
initial-state-cone. The correction is a rotation about the $y-$axis,
which would rotate the post-measurement states as close as possible
to the initial states. This analysis simply consider the rotations
of the axes of the Bloch sphere, to have a good performance, the
length of the post-measurement state should be taken into account,
this makes the optimization of $\beta$ complicated.

It is worth emphasizing that the angle rotated of our initial states
is $\phi$.   One may suspect that when the measurement cancels this
rotation and send the states back to the $x-z$ plane, i.e.,
$\beta=\phi$, the optimal performance can be obtained. This
intuition comes from the optimality proof in Ref.\cite{15}, however
this is not true  as we shall show below.

{By using the average fidelity
$\overline{F'}(\theta,p,\chi,\eta,\phi,\beta)$ in
Eq.(\ref{fertility}), we can calculate
$\frac{\partial\overline{F'}}{\partial \beta}.$ From
$\frac{\partial\overline{F'}}{\partial \beta}|_{\beta=\beta_c}=0,$
$\beta_c$ follows, which maximize the average fidelity
$\overline{F'}$ and takes,
$$\tan\beta_c=-\frac12\frac{\sin\eta\sin^2\theta\sin2\phi}
{\cos\eta\cos^2\theta+\sin^2\theta\sin^2\phi}.$$}   {Clearly, the
$\beta$ that maximize the performance depends not only on $\phi$ and
$\theta$, but also on $\eta$, namely, it connects closely with the
correction $Y_{\pm\eta}$. When $\phi=0$, $\beta_c=0$, returning back
to the earlier scheme.  This observation can be understood as
follows. We denote $U$ the rotation about the $x-$axis, which sends
the initial state back to the $xz-$plane, i.e.,
$\rho_{\pm}=U\tilde{\rho}_{\pm}U^{\dagger}.$ Here,
$\tilde{\rho}_{\pm}=\rho_{\pm}|_{\phi=0}$. Then the resulting state
$\mathcal{C}(\rho^{\prime})$ can be written as,
\begin{equation}
\mathcal{C}(\rho^{\prime})=U\left(
\tilde{Y}_{+\eta}\tilde{M}^{\prime}_{+}
\tilde{\rho}^{\prime}\tilde{M}_{+}^{\prime\dagger}\tilde{Y}_{+\eta}^{\dagger}+
\tilde{Y}_{-\eta}\tilde{M}^{\prime}_{-}
\tilde{\rho}^{\prime}\tilde{M}_{-}^{\prime\dagger}\tilde{Y}_{-\eta}^{\dagger}\right
) U^{\dagger},\label{Cnew}
\end{equation}
where $\tilde{\rho}^{\prime}=(1-p)\tilde{\rho}
+p\tilde{Z}\tilde{\rho}\tilde{Z},$ and
$\tilde{(...)}=U^{\dagger}(...)U.$ This suggests that when the
initial states are written as the same as that in the earlier
scheme, the noise, measurement and the correction all need to
change. Since $X$, $Y$ and $Z$ do not commute with each other, these
changes are not trivial.} We should emphasize that the effect of the
noise given in Eq. (\ref{noise}) is to spoil the off-diagonal
elements of the density matrix, or to shorten the $x-$ and
$y-$component of the Bloch vector for any state, not only for the
states lie in the $xz-$plane, so the aim of our scheme is to protect
states against the same noise as that in the earlier scheme.

In  conclusion, we introduce new measurements to better  the state
protection for a qubit.  The average fidelity is calculated and
discussed. {Numerical optimizations over these parameters show that
the new measurements can extend  the state protection scheme from
special states to general states. This scheme works for a wide range
of initial states and generalize the scheme in the earlier works.}
The construction of the new proposal has several advantages. First,
the initial states are more general, namely the corresponding Bloch
vectors are allowed to lie outside the $xz-$plane, this extends the
range of state protection and makes the scheme more realistic. The
effect of the noise is to shorten the $x$ and $y$ components of the
Bloch sphere, hence the noise is of dephasing. Second, we made use
of a measurement which allow us to mix the $x-$ and $y-$components
of the Bloch sphere, offering a room to improve the performance of
the state protection. Finally, we note that the key elements to our
scheme have already been experimentally demonstrated \cite{14}, we
expect that this extension  of the earlier quantum control scheme
is within reach of current technologies. \ \ \\
\ \ \\
This work is supported by the NSF of China under Grants Nos
61078011,  10935010 and 11175032.

\end{document}